\documentstyle[12pt,epsfig]{article}

\newcommand\sss{\scriptscriptstyle}
\newcommand\as{\mbox{$\alpha_{\sss S}$}}
\newcommand\xt{\mbox{$x_{\sss T}$}}  
\newcommand\et{\mbox{$E_{\sss T}$}}  
  
\newcommand\mur{\mbox{$\mu_{\sss R}$}}  
\newcommand\muf{\mbox{$\mu_{\sss F}$}}  
\def\be{\begin{equation}}         
\def\ee{\end{equation}}
\def\ba{\begin{eqnarray}}
\def\ea{\end{eqnarray}}

\def    \=              {\;=\;}
\def    \frac           #1#2{{#1 \over #2}}

\newcommand{\ccaption}[2]{
  \begin{center}
    \parbox{0.85\textwidth}{
      \caption[#1]{\small\it {#2}}}
  \end{center}    }
\begin{document}
\begin{titlepage}
\nopagebreak
{\flushright{
        \begin{minipage}{4cm}
        CERN-TH/96-278\\
        hep-ph/9610234\\
        \end{minipage}        }

}
\vfill                        
\begin{center}
{\LARGE { \bf \sc Topics in Jet Physics}}\footnote{To appear in the Proceedings of the XIth
Topical Workshop on Hadron Collider Physics, Abano Terme, Padova, Italy, May
1996.}
\vfill                                
\vskip .5cm                               
{\bf Michelangelo L. MANGANO,}
\footnote{On leave of absence from INFN, Pisa, Italy}                           
\vskip .3cm                                      
{CERN, TH Division, Geneva, Switzerland}\\
{\tt mlm@vxcern.cern.ch}                  
\end{center}                                                      
\nopagebreak
\vfill                                
\begin{abstract}
We review two subjects in the theoretical study of jet production at the
Tevatron collider:  the uncertainties in the determination of the partonic
densities inside the proton and the uncertainties in the 
calculation of higher-order corrections to the QCD matrix elements.
\end{abstract}                                                               
\vskip 1cm
CERN-TH/96-278 \hfill \\
October 1996 \hfill
\vfill       
\end{titlepage}


\section{Introduction}\label{sec:intro}
The accurate measurements of inclusive jet distributions recently reported by
the experiments at the Fermilab Tevatron collider~\cite{Abe96,Abachi96}
pose a serious challenge to  the theorists' capability to evaluate the relative
cross sections with matching precision. The current data, and in particular the
discrepancy between data and theory at high \et\ reported by
CDF~\cite{Abe96}, call for an improved assessment of the 
theoretical uncertainties present in the evaluation of the 1-jet inclusive 
\et\ distributions. Several groups have undertaken this task, contributing in
the past several months to a significant progress in the field. 
The duty assigned to me by the organizers of the Workshop was to review this
progress. 

We can divide the problem into two main subjects:  {\it (i)} the study of the
uncertainties in the determination of the partonic densities inside the proton
(discussed here in Section~2), and {\it (ii)} the study of the uncertainties in
the perturbative calculation of the QCD matrix elements (discussed in
Section~3). I will skip all preliminaries on jet definitions and  technical
aspects of the evaluation of the NLO matrix elements, assuming that the reader
is familiar with them. For a complete discussion of
these points the reader is referred to the references quoted below.
                                                                      
\section{Parton Densities}                                
The cross section for 1-jet inclusive production is known today up to NLO
accuracy~\cite{Ellis86,nlojets,jetrad}. It is very unlikely that full NNLO
results will become available in the short term, due to the complexity of the 
calculations. A standard way to estimate the possible size of higher-order
corrections is to evaluate the dependence of the NLO result  w.r.t. variations
of the factorization and renormalization scales. The analysis in the range
$50<\et ({\rm GeV}) <400$  shows a scale dependence of the order of
10--20\%~\cite{nlojets,Giele95}.  This uncertainty is  similar to 
the current experimental uncertainties, and its small size makes the
comparison between theory and data a compelling test of perturbative QCD, as
well as an important probe of the presence of possible new physics.
                                                                   
However accurate the perturbative
evaluation of the partonic matrix elements, the ultimate precision of the
theoretical calculations for the jet cross sections is limited by the
uncertainty in the knowledge of the parton-density functions (PDFs) of the
proton. PDF parametrizations are extracted from global fits to a large variety
of data for which NLO QCD  predictions are available~\cite{Owens92}.  The most
significant inputs are given by  deep inelastic scattering (DIS), Drell--Yan
(DY) and prompt-photon fixed-target production data. DIS provides a direct
probe of the quark densities via the measurement of $F_2(x,Q^2)$. It also gives
a value of \as\ via the scaling violations of $F_2$ at intermediate values of
$x$ ($x\sim 0.3$--$0.5$). In this $x$ range $d\log F_2/d\log Q^2$ receives no
significant contribution from the gluon density 
and $Q^2$ is still sufficiently large for higher-twist effects to be
under control. The contributions of the different quark flavours can be
separated by using a mixture of DIS and DY data for different beams and
targets. The gluon density at small $x$ can be extracted from the scaling
violations of $F_2(x,Q^2)$ (for $x$ values sufficiently small $dF_2/d\log
Q^2\propto \as(Q^2)G(x,Q^2)$). $G(x,Q^2)$ is rather well known today in the
region $10^{-4}<x<0.1$. The most direct probe on the gluon density at larger
values of $x$ so far has been the measurement of large-\et\ prompt photons
produced in fixed-target experiments with proton beams, where the Compton
channel $qg\to q\gamma$ dominates at large $x$ over the quark-sea-suppressed
annihilation channel $q\bar q\to g\gamma$.

\begin{figure}                                               
\centerline{\epsfig{figure=cteqm.seps,width=0.7\textwidth,clip=}}
\ccaption{}{ \label{fig:cteqm}         
Comparison between NLO QCD predictions for the 1-jet inclusive \et\
distribution and Tevatron data (figure taken from ref.~\cite{Lai96}). The
PDF sets used belong to the CTEQ4 family~\cite{Lai96}.}        
\end{figure}                             
The use of the NLO QCD matrix elements, together with one of 
the most recent global PDF fits (CTEQ4~\cite{Lai96}),
leads to the comparison of the theoretical prediction
with the Tevatron data shown in fig.~\ref{fig:cteqm}. The surplus rate 
found in the CDF data~\cite{Abe96}                  
(not confirmed by, but nevertheless consistent within uncertainties with the D0
measurement~\cite{Abachi96}) significantly exceeds the expected theoretical
uncertainty, estimated with the study of the scale dependence. Similar
conclusions are obtained using independent PDF fits, such as those of the MRS
group~\cite{Martin95,Martin95a}.

Several studies have been carried out to establish to which extent
the available data used for the extraction of the PDFs set compelling
constraints on the highest-\et\ tail of the jet distribution. Before
illustrating the results of these studies, I will present a few remarks.
                                                                      
\begin{figure}                                               
\centerline{\epsfig{figure=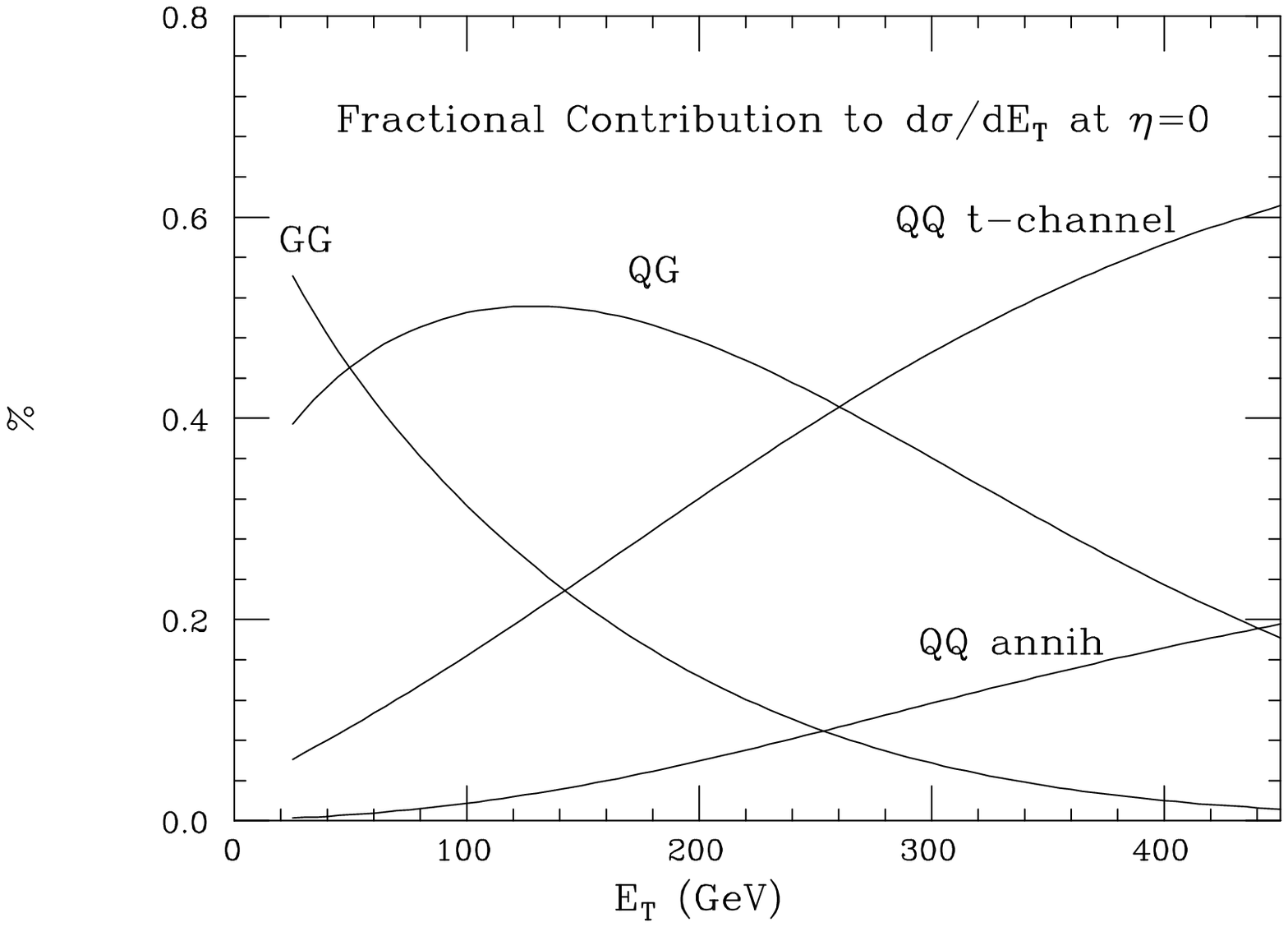,width=0.49\textwidth,clip=}
            \epsfig{figure=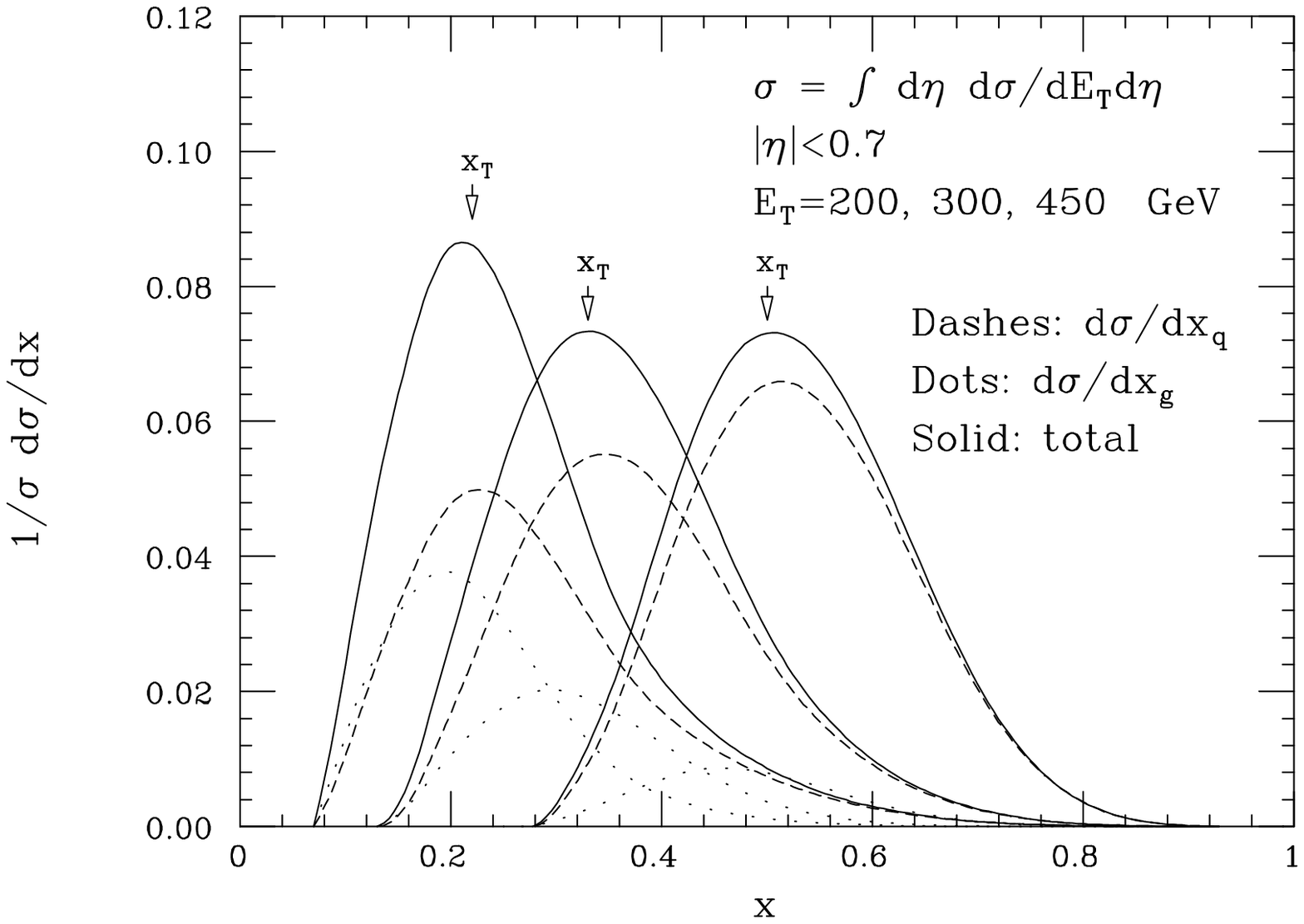,width=0.49\textwidth,clip=}}
\ccaption{}{ \label{fig:isfrac}                 
Left: relative contribution of the different initial states for the production
of a jet at $\eta=0$ as a function of the jet \et. Right: distribution in $x$
for the production of jets with \et=200, 300 and 450~GeV.}
\end{figure}                          

The values of $x$ that are probed and the flavour content of the initial state 
in jet production depend on the jet \et. In fig.~\ref{fig:isfrac}{\it a}
I show the
fractional contribution to $d\sigma/d\et$ given by the three possible 
intiial-state channels ($gg$, $qg+\bar q g$ and $qq+q\bar q$). The $q\bar q$
channel has been separated into the contributions of processes which only admit
$s$-channel exchange (namely $q\bar q\to gg$ and $q_i\bar q_i \to q_j\bar
q_j$), labelled as ``QQ annih'', and those which also admit a $t$-channel
exchange (e.g. $q_i\bar q_i \to q_i\bar q_i$ and $q_i q_j \to q_i q_j$),
labelled as ``QQ $t$-channel''. These last processes are seen from the figure to
dominate the production at the largest \et\ values observed at the Tevatron,
namely $\et=450$~GeV.

The ranges of $x$ that are relevant to the production of jets at three 
different energies are shown in fig.~\ref{fig:isfrac}{\it b}. 
While the distributions
as a function of $x$ peak at the obvious value $x=\xt\equiv
2\et/\sqrt{S}$, long tails are present even at $x$
values significantly larger than \xt. As a result, in order to have accurate
predictions for jet production at the highest transverse energies observed at
the Tevatron, $\et\sim 450$~GeV, the PDFs should be known for values of
$x$ up to about 0.6.

\begin{figure}                                               
\centerline{\epsfig{figure=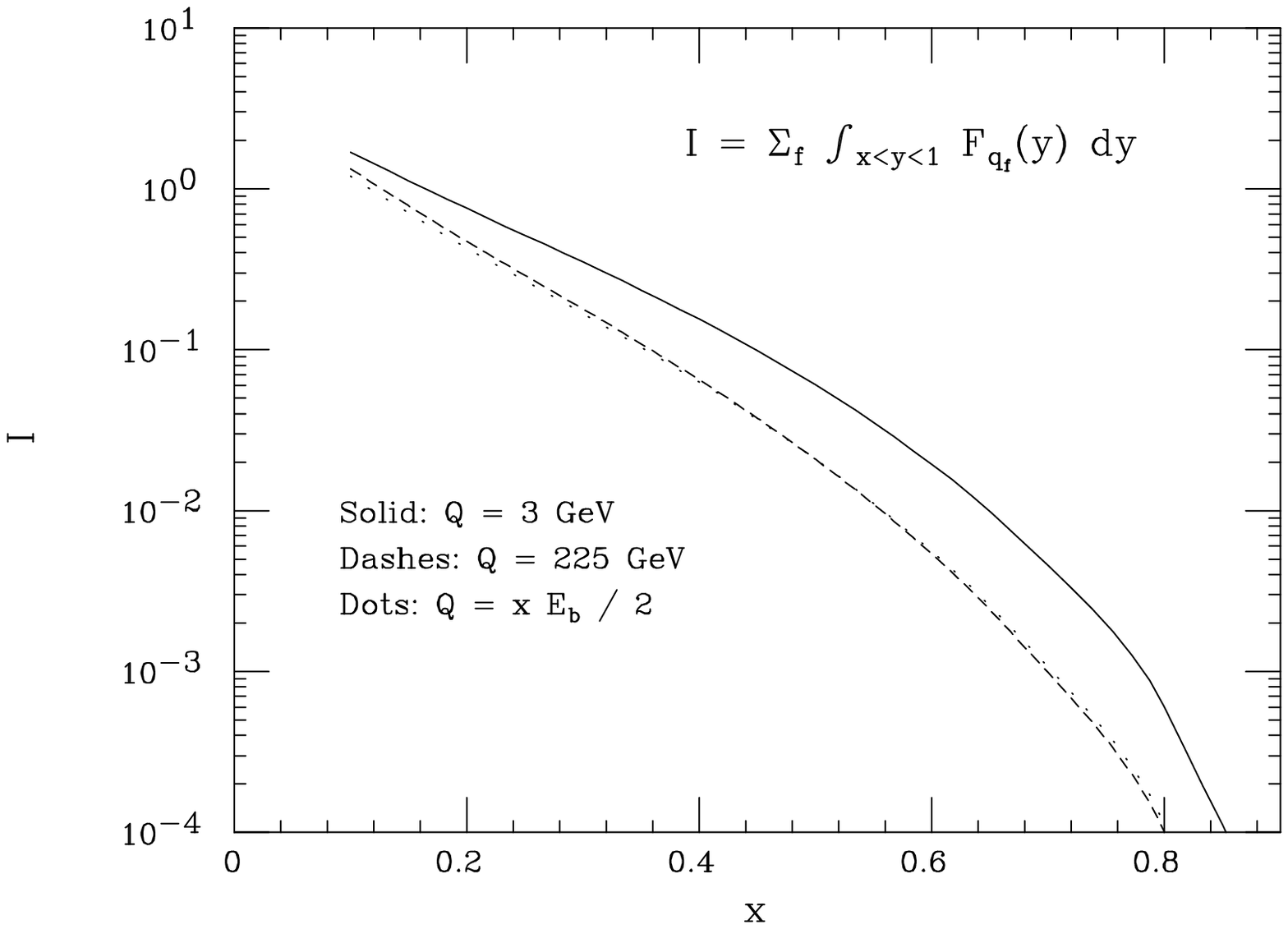,width=0.7\textwidth,clip=}}
\ccaption{}{ \label{fig:pdfevo}         
Evolution of the integrated quark density, summed over all $q$ and $\bar q$
flavours.}                             
\end{figure}

Since the dominant contribution to the jet rate at 450~GeV is given by quark
initial states, it is natural to explore the possibility that the quark
densities are not known with sufficient accuracy. First of all, given that the
quark densities at large $x$ are extracted from DIS data at low $Q^2$, it is
important to establish what is the range in $x$ at low $Q^2$ of
relevance for the jet rates at high $Q^2$. In fact, given the QCD perturbative
evolution, it is natural to expect that most of the quarks found at large $Q^2$
with $x$ of the order of 0.5 will actually come from quarks having a much
larger value of $x$ at the lowest $Q^2$ scales explored by the DIS experiments. 
Figure~\ref{fig:pdfevo} gives the integral of the quark density above a given
value of $x$ at $Q=3$~GeV (a typical scale for DIS data) and at $Q=225$~GeV
(the scale for production of a 450~GeV jet, when using the standard choice
$\mur=\muf=\et/2$). 

For $x$ values in the range 0.5--0.6, the evolution reduces by a factor of 3
the probability to find a quark. This in itself is a dramatic               
change, which prompts some observations. To start with,
it is important to point out that the size of the depletion at
large $x$ is controlled by the value of \as, since at large $x$
$d\log F_2/d\log Q^2\propto -\as(Q^2)$. The larger the value of \as, the
stronger the depletion. Therefore, although one might expect that a
larger value of \as\ would generate an increase in the rate of jet production,
the opposite is actually true at large \et. This is shown, for example, in
fig.~\ref{fig:mrsf1}{\it a}, which is taken from the work of ref.~\cite{Glover96}.
The fit labelled $J$ (dashed line) is obtained by including the CDF jet data
in the global PDF fit. In order to accommodate the shape of the lower-energy jet
data, the fit selects a value of \as\ slightly larger than that preferred by
the pure DIS analysis ($\as(M_Z^2)=0.120$ vs. $\as(M_Z^2)=0.113$)\footnote{It is
important to notice that this result is consistent with an independent     
evaluation of \as\ from the CDF jet data~\cite{Giele96}.}.
The larger                                                
value of \as\ is reflected in the lower rate predicted by this set of PDFs at
high \et. As an additional comment, it should be remarked that such a
significant change in the quark density, induced by the evolution, might be
sensitive to the presence of large higher-order corrections to the evolution
equations. In the usual analyses, the evolution is performed at NLO,
consistently with the level of accuracy of the matrix element evaluation. In
the region of large $x$ and low $Q^2$, however, big corrections could come from
the resummation of threshold effects, as suggested in~\cite{Catani96a}. The
inclusion of these effects has never been carried out, and it is hard to
evaluate at this stage what their impact could be.
                                                                       
\begin{figure}                                               
\centerline{\epsfig{figure=mrsf1.seps,width=0.49\textwidth,clip=}
            \epsfig{figure=mrsf2.seps,width=0.49\textwidth,clip=}}
\ccaption{}{ \label{fig:mrsf1}                  
Left:
Comparison between NLO QCD predictions for the 1-jet inclusive \et\
distribution and CDF data (figure taken from ref.~\cite{Glover96}). The
dashed line uses a PDF fit obtained including the CDF jet data in the fit. The
dash-dotted line uses a PDF fit which includes 
the CDF data but leaves out all DIS data for $x>0.2$.
Right:                                               
Comparison between the PDF fits used in the left figure
and BCDMS data for                   
$F_2$ (figure taken from ref.~\cite{Glover96}).}                             
\end{figure}                                  

   
Another interesting aspect of the evolution displayed in
fig.~\ref{fig:pdfevo} is that the number of quarks with $x>0.6$ at $Q=225$~GeV
is approximately equal to the number of quarks with $x>0.7$ at $Q=3$~GeV. Since
the quark density drops very quickly (almost one order of magnitude in going
from $x=0.7$ to $x=0.8$), we can conclude that quarks with $x>0.7$ at $Q=3$~GeV
have no significant impact on the density of quarks in the region $x<0.6$ at
$Q\sim 200$~GeV. Therefore provided the low-energy DIS data are solid in the
range $x<0.7$, we expect no significant uncertainty  due to the knowledge of
the quark PDFs. The data are indeed quite accurate in this region, as shown in
fig.~\ref{fig:mrsf1}{\it b}. For this reason, when the jet data are included in a 
global fit of PDFs the high-\et\ points have no impact on the quark densities,
which are almost entirely controlled by the small error of the DIS data points.

Figure~\ref{fig:mrsf1}{\it a}
 displays also the results of a fit obtained by removing
all DIS data in the region $x>0.2$ while including the CDF jet data (fit
$J^{\prime}$). The absence of independent constraints on the quark densities
allows the jet data to drive them arbitrarily high, and an acceptable fit of
the jet data is possible. However, when this fit is compared with 
the existing DIS data, the disagreement is absolutely unacceptable. The
prediction for $F_2$ obtained by the $J^{\prime}$ fit is shown in
fig.~\ref{fig:mrsf1}{\it b}, compared with the data and with the results of
standard fits. At $x=0.55$, $F_2$ is increased by about 20--30\%, which is what
one would in fact expect in order to accommodate the 50\% excess rate observed
by CDF. The authors of ref.~\cite{Glover96} therefore conclude that it is not
possible to fudge the quark densities in order to fit the CDF jet data without
an unacceptable clash with the high-precision DIS data.
                                                       
\begin{figure}                                               
\centerline{\epsfig{figure=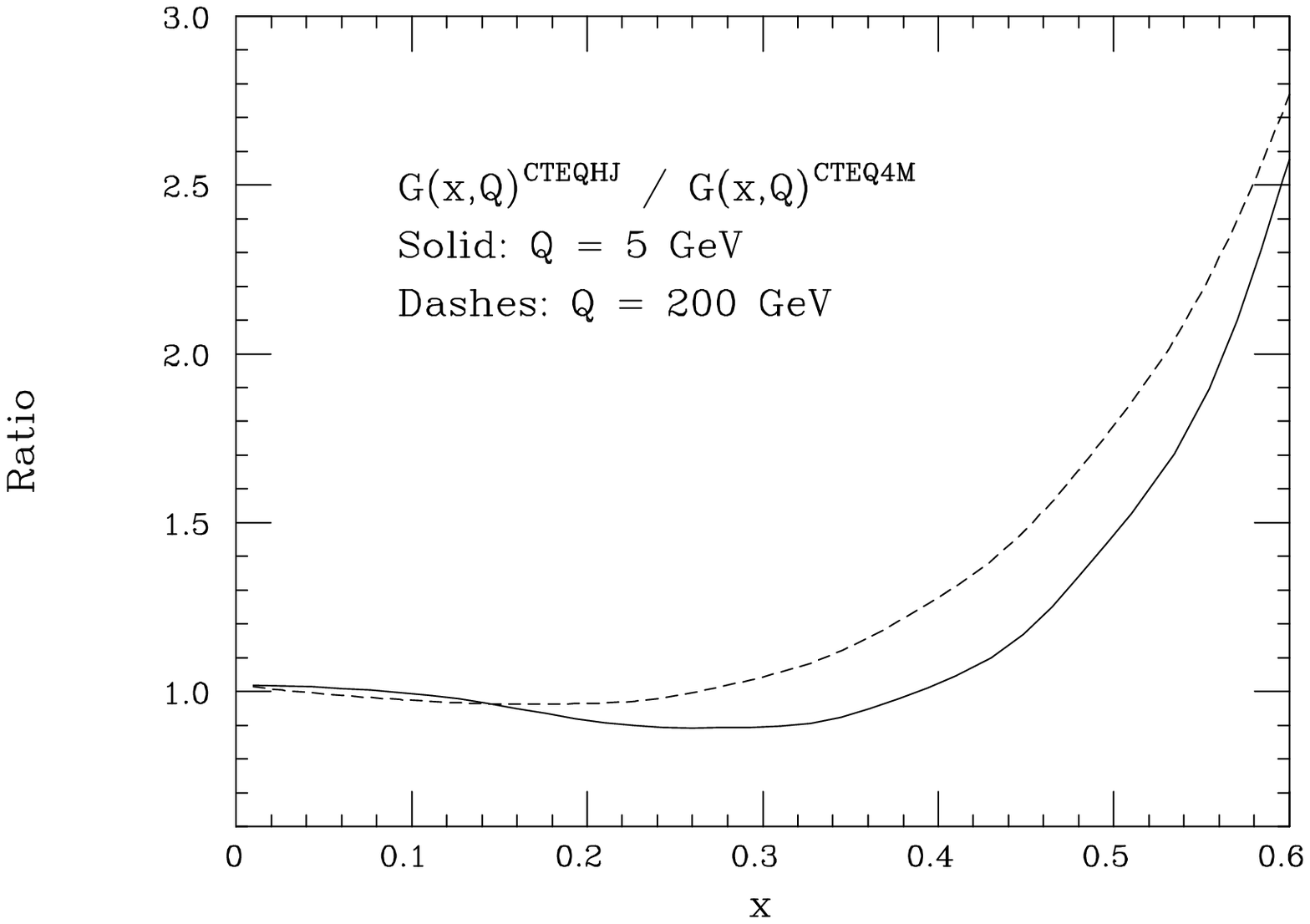,width=0.49\textwidth,clip=}
            \epsfig{figure=cteqhj.seps,width=0.49\textwidth,clip=}}
\ccaption{}{ \label{fig:cteqrat}                 
Left: ratio of the CTEQHJ and CTEQ4M gluon densities. Right: 
comparison between NLO QCD predictions for the 1-jet inclusive \et\
distribution using the CTEQHJ set 
and the Tevatron data (figure taken from ref.~\cite{Lai96}).}
\end{figure}                          

An alternative approach was pursued by the CTEQ group. These authors exploited
the fact that the experimental and theoretical uncertainties present in the
analysis of the prompt photon data might leave enough room to accommodate larger
gluon densities at large $x$. From fig.~\ref{fig:isfrac} we see that in order
for processes with initial-state gluons to explain the 50\% excess rate, the
gluon density should increase by at least a factor of 2. The CTEQ group
performed a global fit~\cite{Huston96,Lai96}
including the high-\et\ CDF data and assigning to them a
larger weight in the $\chi^2$ determination. The gluon density obtained in this
way, called CTEQHJ, is compared with a ``standard'' gluon density in 
fig.~\ref{fig:cteqrat}{\it a}. The
global $\chi^2$ of the fit is marginally worse than the best $\chi^2$ (a
detailed discussion of this point can be found in ref.~\cite{Lai96}).
In particular, the theoretical predictions for the
prompt-photon spectrum obtained using this set of PDFs are perfectly
consistent, within uncertainties, with the experimental data~\cite{Huston96}.
The comparison between QCD and data for the jet-\et\ distributions is shown in
fig.~\ref{fig:cteqrat}{\it b}. 
As can be seen, most of the excess rate has been removed by
this choice of PDFs.

 
Before concluding this section I would like to make a few more comments. During
the summer, new results on neutrino DIS from the CCFR collaboration were
released~\cite{Harris96}.
The most significant result of these analyses is a new extraction of \as\ from
the large-$x$ scaling violations in $F_2$ and $xF_3$. 
Contrary to all previous extractions of \as\ from DIS\cite{Kataev96},
this measurement is in perfect agreement with the values of \as\ extracted at
higher energies (e.g. at LEP) or in $\tau$ decays~\cite{Girone96}.  In
particular, the new CCFR measurement of $\as(M_Z)=0.119\pm0.005$ is
significantly larger than the previous average of DIS measurements,     
$\as(M_Z)=0.112\pm0.004$~\cite{Barnett96}.
It is hard to evaluate what the impact of these data will be on the calculation
of jet cross sections for the Tevatron. The $F_2$ and $xF_3$ 
data have not been released as yet, so that no global fit
of PDFs and their evolution is available.
It is extremely encouraging that the new DIS \as\ value is consistent with
that extracted from the analysis of the jet data~\cite{Giele96}. 
Even more so because these
determinations are now consistent with the extraction of \as\ from 
other independent measurements~\cite{Bethke96}.          
It is likely that, because of
the enhanced evolution of the large-$x$ valence densities, the high-\et\ CDF
anomaly will be even more significant, as we already discussed above. As the
CTEQ studies have shown, however, this could be absorbed into 
an acceptable change in the gluon densities. Independent measurements of the
large-$x$ gluon component of the proton are at this point of extreme
importance. Higher statistics prompt-photon production data are becoming
available~\cite{Zielinski96}, and will hopefully help to settle this issue
soon.
                                                    
\section{Higher orders in perturbation theory}
As mentioned in the previous section, 
the study of the scale dependence in the range
$50<\et ({\rm GeV}) <400$  shows an effect of the order of
10--20\%~\cite{nlojets,Giele95}. It is reasonable to expect that the
NNLO corrections, in this \et\ range, will be of this size. For larger \et\
values the scale dependence becomes more and more
pronounced~\cite{nlojets,Giele95}, indicating that NNLO effects can no longer
be neglected. Whether the actual threshold for the onset of significant
NNLO contributions is 400, 450 or 600 GeV, is clearly a question of
extreme relevance for the interpretation of the  CDF data.
                                                          
To answer this question we must find a more solid way of estimating the 
higher-order corrections than just looking at the scale dependence of the 
lower-order results. Direct estimates of higher-order corrections are possible
when these are dominated by terms whose structure  does not depend on the
specific details of the hard process under consideration.  A typical example is
given by the large logarithms that are associated to the emission of collinear
gluons from the initial state of a hard process. These logarithms are
independent of the details of the hard process, and can be resummed to all
orders of perturbation theory, giving rise to the standard DGLAP equation which
governs the evolution of the parton densities inside the proton. 
                                                                                
The resummation of higher-order contributions is generally possible when two
different hard scales (say $\mu$ and $Q$, with $Q\gg\mu$) 
are present in the problem. Gluons whose wavelength is of the order of $1/\mu$
cannot be sensitive to phenomena taking place at distances of the order of
$1/Q$. If their emission gives rise to large corrections of order
$\as\log^k(Q/\mu)$ ($k\le 2$), as is often the case, these corrections are
therefore universal and can be resummed (for a thorough discussion of these
issues at NLO, and for a complete set of references to the relevant literature,
see refs.~\cite{Sterman,CataniTrentadue}).
In the case of high-energy jet           
production, the two different hard scales correspond to the energy of the jet
itself and  the maximum amount of energy that can be released by initial-state
radiation. The closer one gets to the threshold $M^2_{jj}\to S_{had}$ (where
$M_{jj}$ is the invariant mass of the dijet system and $\sqrt{S_{had}}$ is the
total energy in the $p\bar p$ CoM frame), the smaller the amount of energy
available for radiation. Given that parton densities fall steeply, the
phase space available for soft radiation is already significantly reduced
before $\tau=M^2_{jj}/S_{had}$  approaches 1. In the case of DY production in
fixed-target experiments, for example, large corrections are already found for
values of $\tau$ of the order of 0.5~\cite{Curci81,Appel88}.  
It is believed that higher powers of                      
$\as\log^2(1-\tau)$ are responsible for the increased sensitivity of the jet
cross section at high \et, and that these are the leading 
corrections that need to be resummed to provide a better estimate of
the jet cross section beyond NLO.                                    

A technical discussion of how the resummation of threshold corrections is
performed, in addition to references to the original literature, can be found
in ref.~\cite{Catani96a}.
Here I will limit myself to presenting one result of interest for the study of
high-energy jet production. Rather than considering the resummation corrections
to the \et\ distribution of jets, I will consider the case of high-mass dijet
pairs. The reason is that resummation is simpler in this case, since for this
variable all Sudakov effects are present only in the evolution of the initial
state, so that the formalism is almost identical to that of the DY processes.
The study of the resummation for the 1-jet inclusive \et\ distribution is in
progress.

\begin{figure}                             
\centerline{\epsfig{figure=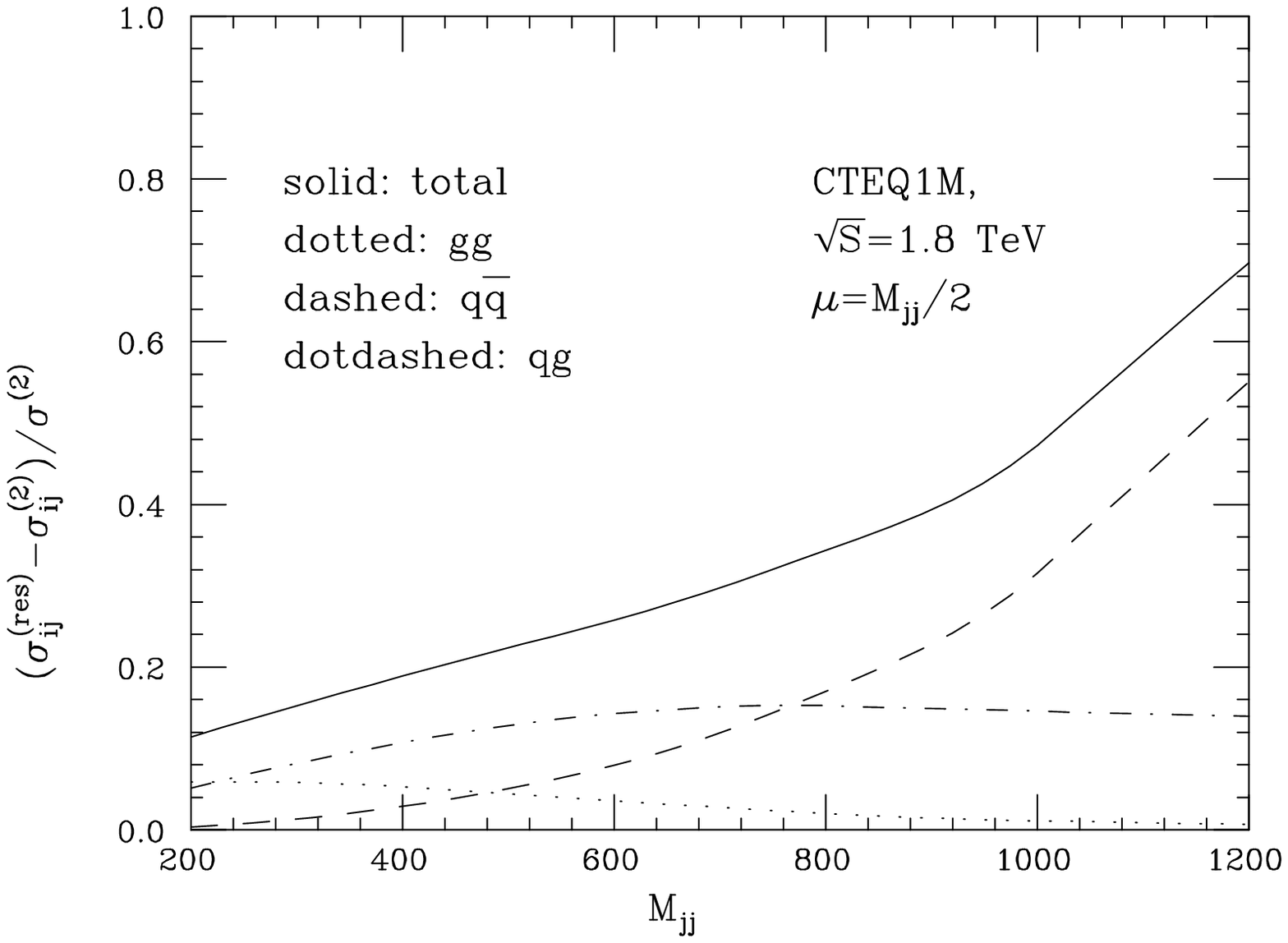,width=0.66\textwidth,clip=}}
\ccaption{}{ \label{jetcteq1}
Contribution of gluon resummation at order $\as^3$ and higher, relative to the
LO result, for the invariant mass distribution of jet pairs at the Tevatron.}
\vfill
\centerline{\epsfig{figure=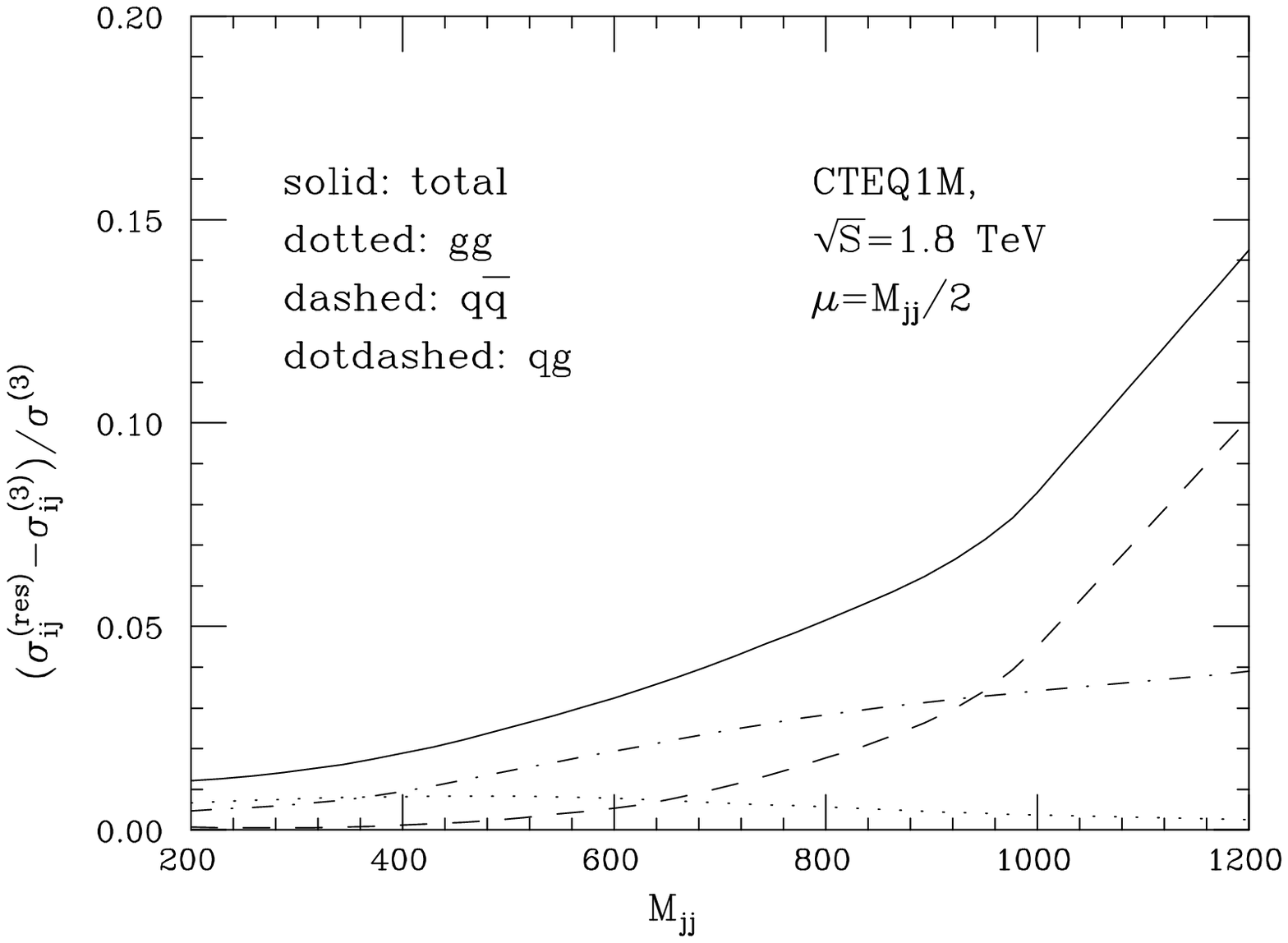,width=0.66\textwidth,clip=}}
\ccaption{}{ \label{jetcteq2}
Contribution of gluon resummation at order $\as^4$ and higher, relative to the
truncated ${\cal O}(\as^3)$ result,
for the invariant mass distribution of jet pairs at the Tevatron.}
\end{figure}

In fig.~\ref{jetcteq1} I show the following quantities
\be                                         
\frac{\delta^{(3)}_{\rm gg}}{\sigma^{(2)}}\,,\quad
\frac{\delta^{(3)}_{\rm qg}}{\sigma^{(2)}}\,,\quad
\frac{\delta^{(3)}_{\rm q\bar{q}}}{\sigma^{(2)}}\,,\quad
\frac{\delta^{(3)}_{\rm gg}+\delta^{(3)}_{\rm qg}+\delta^{(3)}_{\rm
q\bar{q}}}{\sigma^{(2)}} \, ,
\ee
where $\delta^{(3)}$ is equal to
the resummed hadronic cross section
in which the terms of order $\as^2$
have been subtracted, and
$\sigma^{(2)}=\sigma^{(2)}_{\rm gg}+\sigma^{(2)}_{\rm qg}
 +\sigma^{(2)}_{\rm q\bar{q}}$
is the full hadronic LO cross section
(of order $\as^2$).
I use as a reference renormalization and factorization
scale $\mu=M_{jj}/2$. Notice that for
large invariant masses the effects of higher orders are large.
To understand how much is due to the first-order corrections
(which are exactly calculable \cite{Ellis92,jetrad})
and how much to corrections of order       
$\as^4$ and higher, I show the following quantities in fig.~\ref{jetcteq2}:
\be                   
\frac{\delta^{(4)}_{\rm gg}}{\sigma^{(3)}}\,,\quad
\frac{\delta^{(4)}_{\rm qg}}{\sigma^{(3)}}\,,\quad \frac{\delta^{(4)}_{\rm
    q\bar{q}}}{\sigma^{(3)}}\,,\quad \frac{\delta^{(4)}_{\rm
    gg}+\delta^{(4)}_{\rm qg}+\delta^{(4)}_{\rm q\bar{q}}}{\sigma^{(3)}} \; ,
\ee
where $\delta^{(4)}$ is now equal to the resummed hadronic cross section
with terms of order $\as^3$ subtracted, and $\sigma^{(3)}$ is an approximation
to the full NL cross section, summed over all subprocesses, obtained by
truncating the resummation formula at order $\as^3$.  This figure shows that
most of the large $K$ factor is due to the pure NLO corrections, with
the resummation of higher-order soft-gluon effects contributing only an
additional 10\% at dijet masses of the order of 1~TeV.  
Only above $M_{jj}>1200$~GeV, corresponding approximately to $\et>600$~GeV, are
the resummation effects non-negligible.

These results
should only be taken as an indication of the order of magnitude of the
correction, since we have not included here a study of the resummation
effects on the determination of the parton densities.  
As was already mentioned previously, it is quite possible that
resummation effects significantly influence the determination of the large-$x$
structure functions from low energy data. Equally important effects could
appear in the evolution from low to high $Q^2$. 
It would therefore be important to       
reexamine the extraction of the large-$x$ non-singlet structure functions, in
the light of the resummation results for the DIS process, before firmer
conclusions can be drawn on the significance of the present jet cross section
discrepancy.

From this preliminary study it seems, however, unlikely that the full 30--50\%
excess reported by CDF for jet \et's in the range 300--450~GeV could be    
explained by Sudakov resummation effects in the hard process. 

Before concluding this section, I want to mention a puzzling observation
recently made in ref.~\cite{Kramer96}. The authors studied the scheme
dependence of the jet cross section, by using the $\overline{\rm MS}$ and DIS
versions of the same PDF fit, folded with the NLO jet cross section evaluated
in the two respective schemes. The two evaluations should yield equal
results, up to terms of NNLO. However, the differences turn out to be
numerically large, of the order of 40\% at the highest \et's. The DIS
calculation, in particular, is in good agreement with the CDF data. Whether this
difference corresponds to a genuine scheme dependence, and therefore a true
uncertainty due to the ignorance on corrections beyond NLO, or whether it is an
artefact of the way the DIS PDFs were extracted or evolved, this is still not
clear. 
                        
\section{Conclusions}                
The main conclusions of this review can be summarized as follows:
\begin{itemize}
\item The NLO approximation is solid up to $\et=500$~GeV, with uncertainties at
the level of 10--20\%. The effects of higher-order corrections, estimated here
in the case of dijet production by resumming the largest leading soft
logarithms appearing as $x\to 1$, will become non-negligible only for
transverse energies larger than 600~GeV.
\item Modifications, at large $x$, of the gluon density evaluated so far seem
possible without affecting the overall quality of global PDF fits. It seems
possible to absorb a large fraction of the excess of high-\et\ jets reported by
CDF, by allowing for a harder gluon at large $x$, without affecting the
comparison of theory with other sets of data.
\item While resummation corrections do not seem important in the evaluation of
the hard cross sections, their effect could be non-negligible on the evolution
of the valence quark densities from the low-$Q^2$ region (where they are
measured) to the high-$Q^2$ region relevant for the Tevatron jets.
\item The new measurements of $F_2$ and $xF_3$ by CCFR call for new analyses of
the global fits to the PDFs. Only at that point  will a new assessment of the
situation with the high-\et\ Tevatron jets be possible.
\end{itemize}
The last two points indicate that, in spite of the significant
progress we witnessed recently, there is still room for new ideas and for
improvement.

\end{document}